\documentstyle[twocolumn,prl,aps,psfig,floats,amssymb]{revtex}

\begin{document} 
\draft

\twocolumn[
\hsize\textwidth\columnwidth\hsize\csname @twocolumnfalse\endcsname

\title{Smoothed universal correlations in the two-dimensional Anderson
model}

\author{Ville Uski$^1$, Bernhard Mehlig$^{2}$\cite{bm-addr}, Rudolf A.
  R\"omer$^1$, and Michael Schreiber$^1$}

\address{ $^1$Institut f\"ur Physik, Technische Universit\"at, D-09107
  Chemnitz, Germany \\ $^2$Theoretical Physics, University of Oxford,
  1 Keble Road, Oxford, UK}

\maketitle

\begin{abstract}
  We report on calculations of {\em smoothed}\ spectral correlations
  in the two-dimensional Anderson model for weak disorder.  As pointed
  out in (M.\ Wilkinson, J.\ Phys.\ A: Math.\ Gen.\ {\bf 21}, 1173
  (1988)), an analysis of the smoothing dependence of the correlation
  functions provides a sensitive means of establishing consistency
  with random matrix theory. We use a semiclassical approach to
  describe these fluctuations and offer a detailed comparison between
  numerical and analytical calculations for an exhaustive set of
  two-point correlation functions.  We consider parametric correlation
  functions with an external Aharonov-Bohm flux as a parameter and
  discuss two cases, namely broken time-reversal invariance and
  partial breaking of time-reversal invariance.  Three types of
  correlation functions are considered: density-of-states, velocity
  and matrix element correlation functions. For the values of
  smoothing parameter close to the mean level spacing the
  semiclassical expressions and the numerical results agree quite well
  in the whole range of the magnetic flux.
\end{abstract}


\narrowtext 
]

\section{Introduction}
\label{sec:intro}

Disordered quantum systems in the metallic regime exhibit irregular
fluctuations of eigenvalues\cite{boh89}, eigenfunctions\cite{prig95}
and also of matrix elements\cite{tan95,meh98c}.  
Parametric fluctuations have been
discussed in \cite{others}.
In the metallic regime, which is characterized by a large conductance
$g\gg 1$, such fluctuations can be described by random matrix theory
(RMT) \cite{wig51} on energy scales smaller 
than the Thouless energy
$E_{\rm D} = g \Delta$, 
($\Delta$ is the mean level spacing).
Alternatively, semiclassical methods may be used in this regime,
as suggested in \cite{wilk88}. 
A semiclassical estimate for parametric
correlations of level velocities is given in \cite{ber94}.
Matrix element correlations
are discussed in \cite{meh98c,wilk96,kea98}.
Within a semiclassical approach it is essential to
incorporate level broadening and work with
smoothed correlation functions.
The level broadening $\epsilon$ needs to be
larger or of the order of
the mean level spacing, $\epsilon \gtrsim \Delta$.
This ensures that the periodic orbit sums are truncated
in such a way that only orbits with periods $T_p$
shorter than the Heisenberg time $t_{\rm H} = 2\pi\hbar/\Delta$
contribute.  The results reported in
\cite{ber94,wilk96,eck95} predict characteristic dependences
on the smoothing.
As pointed out in \cite{wilk88}, the smoothing
dependence of the fluctuations provides a sensitive
means of establishing consistency with RMT.
The semiclassical approach provides a natural
approach of incorporating such a smoothing.

In this paper, we report on extensive numerical calculations
of correlation functions in the two-dimensional (2D) Anderson model
of localization \cite{and58}
in the metallic regime, compare also \cite{sim93a}. 
In the limit of large $g$, the statistical properties of
eigenvalues and eigenvectors in this model
can be described by RMT on energy
scales smaller than the Thouless energy $E_{\rm D} = g\Delta$,
compare \cite{hof93,zha97,mul97}.

We calculate parametric correlation
functions (where an Aharonov-Bohm flux is used as an external
parameter) as well as fluctuations in systems with
weakly broken time-reversal ($\sf T$)-symmetry.
We calculate three types of correlation functions,
namely correlations of the density of states
\cite{sim93a}, of velocities \cite{ber94,sim93a}
and of matrix elements \cite{meh98c,wilk88,wilk96,kea98,eck95}.
$\sf T$-invariance is broken by means of an Aharonov-Bohm flux
$\phi$. According to RMT, fluctuations in a $\sf T$-invariant system, where 
$\phi=0$, follow the statistics of the Gaussian orthogonal ensemble 
(GOE). At $\phi\simeq\phi_0/4$, where $\phi_0 = hc/e$ denotes the flux quantum,
$\sf T$-invariance is fully broken, and RMT predicts the behavior of 
the Gaussian unitary ensemble (GUE).
For $\phi \ll \phi_0/4$ $\sf T$-invariance is only weakly broken.
In this case the correlation functions are described by the 
Pandey-Mehta ensemble \cite{pan83}. The effect of a weak
magnetic field can be exhibited particularly transparently
within a semiclassical approach.

All correlation functions calculated in the following
will be expressed in terms of smoothed spectral densities.
In the literature, Lorentzian \cite{ber94}
as well as Gaussian broadening \cite{wilk88,wilk96}
have been used. For numerical calculations,
Gaussian broadened densities are much more convenient,
since one invariably deals with finite stretches
of spectra, and boundary effects are less pronounced
due to faster decaying tails in the Gaussian case.

We calculate these correlation functions numerically,
analyze the smoothing dependence in detail,
and determine the three non-universal constants,
namely the mean level spacing $\Delta$, 
the conductance $g$ and, in the
case of matrix  element correlations,
the variance $\sigma_{\rm off}^2$ of off-diagonal matrix elements.
We report on successes of and problems with the
semiclassical approach in describing correlations
in the Anderson model in the metallic regime.

The article is organized as follows. In Sec.\ \ref{sec:semiclass}, we
recall those features of the semiclassical approach that will be used
in the derivation of the correlation functions.  In Sec.\ \ref{sec:am}
we describe the Anderson model of localization in the weakly
disordered regime at finite external flux. In Sec.\ \ref{sec:goegue}
we study the correlation functions for the transition from the 
GOE to GUE transition, and
in Sec.\ \ref{sec:para} the parametric correlation functions, and
compare the semiclassical formulae with the results from the numerical
simulations of the Anderson model. In Sec.\ \ref{sec:histo} we study
the distribution functions of our results and compare them to
theoretical predictions.  We conclude in Sec.\ \ref{sec:concl} with a
discussion of our results.

\section{The semiclassical approach to universal correlations}
\label{sec:semiclass}

In this paper, we calculate  correlation functions of the following
densities. We consider the density of states defined as
\begin{equation}
  d(E,\phi) = \sum_\alpha \delta_\epsilon\big[E-E_\alpha(\phi)\big]\,.
\label{eq:dos}
\end{equation}
Here, $E_\alpha(\phi)$ are the quantum eigenvalues and
$ \delta_\epsilon(E) = (\sqrt{2\pi}\epsilon)^{-1/2}\, \exp
  \left(-E^2/2\epsilon^2\right)$.
Second, we consider
the density of parametric velocities \cite{sim93a,wilk89,wilk92b}
\begin{equation}
\label{eq:vdos}
  d_{v}(E,\phi) = \sum_\alpha \frac{\partial E_\alpha}
  {\partial\phi}\, \delta_\epsilon\big[E-E_\alpha(\phi)\big]\,.
\end{equation}
After unfolding (see the next section), the average level velocity is zero.
Third, we compute correlation functions involving a
density of expectation values
\begin{equation}
\label{eq:mdos}
  d_{m}(E,\phi) = \sum_\alpha A_{\alpha\alpha}\,
  \delta_\epsilon[E-E_\alpha(\phi)\big]\,,
\end{equation}
with $A_{\alpha\alpha}= \langle \psi_\alpha(\phi)|\widehat
A|\psi_\alpha(\phi)\rangle$, where $\psi_\alpha(\phi)$ are
the eigenfunctions corresponding to $E_\alpha(\phi)$.
$\widehat A$ is an operator of some real-space observable, not
commuting with the Hamilton operator $\widehat H$.
It is assumed that $\langle A_{\alpha\alpha}\rangle = 0$.

In
all cases, the corresponding densities are
decomposed into a smooth and an
oscillatory part,
\begin{equation}
  d(E,\phi) = \langle d(E,\phi) \rangle + \widetilde d(E,\phi)\,,
\end{equation}
where the first term denotes a mean contribution, and the
second term is a fluctuating part which vanishes
upon disorder averaging. 
The mean parts of the densities
(\ref{eq:vdos}) and (\ref{eq:mdos}) are approximately zero.

For all three densities, we calculate correlation functions of the
type
\begin{equation}
\label{eq:corr}
{\cal C} (\phi_1,\phi_2) = \langle \widetilde d(E,\phi_1)\,
\widetilde{d}^\ast(E,\phi_2)\rangle_E\,.
\end{equation}
The average $\langle \cdots\rangle_E$ denotes an appropriate
average, e.g., over disorder realizations and/or energy in the metallic
regime. 
Semiclassically, such correlation functions
can be calculated using a representation of the densities
in terms of the classical periodic orbits\cite{gut67},
\begin{eqnarray}\label{eq:gutz}
  \lefteqn{\widetilde d(E,\phi) = }\\ & & \frac{1}{2\pi\hbar}
  \sum_{p,r} w_{p,r} \, T_p \, \exp \Big[\!-\!\frac{i}{\hbar} r S_{p}(E)
  +2\pi i r n_p \frac{\phi}{\phi_0} -
  \frac{\epsilon^2r^2T_p^2}{2\hbar^2}\Big]\,.\nonumber
\end{eqnarray}
Here, the sum is over periodic orbits $p$ and their repetitions $r$.
The $w_{p,r}$ are the semiclassical weights, including Maslov indices.
In general they are complex quantities.  $T_p$ denote the periods and
$S_p(E)$ the actions of the periodic orbits $p$. Their windings around
the flux $\phi$
are counted by the winding numbers $n_p$.
Similar expressions can be derived for densities weighted
with level velocities 
or matrix elements 
as shown e.g.  in \cite{wilk88,ber94,eck95,eckXX}.

Correlation functions of the type (\ref{eq:corr}) thus involve
double sums over periodic orbits. It is argued \cite{ber91}
that the average $\langle\cdots\rangle_E$ suppresses
the non-diagonal contributions to this double sum.
This is certainly the case for $\epsilon > \Delta$.
Within the diagonal approximation which amounts to neglecting
the non-diagonal contributions we obtain
\begin{eqnarray}
  {\cal C}(\phi_1, \phi_2)= 
& & \frac{1}{(2\pi\hbar)^2} \sum_{pr} |w_{pr}|^2\,T_p^2 \, 
    {\rm e}^{-\epsilon^2 r^2 T_p^2/\hbar^2}\nonumber \\ 
&&\times  \left(
{\rm e}^{2\pi i\, {n}_p \frac{\phi_1 - \phi_2}{\phi_0}} +
{\rm e}^{2\pi i\, {n}_p \frac{\phi_1 + \phi_2}{\phi_0}}
\right)\,.
\label{eq:diag}
\end{eqnarray}
We can then make use of the sum rule \cite{han84}
\begin{equation}
  \sum_{p} |w_{p}|^2\, T_p^2\, f(T_p) \simeq \int_0^T\!dT\,T\,f(T)\, ,
\label{eq:sumrule}
\end{equation}
which is valid when long periods $T_p$ dominate the sum in
(\ref{eq:diag}). In order to apply (\ref{eq:sumrule}) to
(\ref{eq:diag}), two further approximations are necessary. First,
repetitions are neglected, the usual argument being that periodic
orbits proliferate exponentially. Second, assuming that the
winding numbers are Gaussian distributed, Eq.\ (\ref{eq:diag}) is
averaged over the distribution of winding numbers $P(n,T) = (2\pi
\lambda T)^{-1/2}\,\exp(-n^2/2\lambda T)$ \cite{dit98}. The parameter 
$\lambda=2{\cal D}/L^2$, where $\cal D$ is the
diffusion constant and $L$ the system size.  Evaluating the discrete
average over the winding numbers by Poisson summation, we then obtain
the desired semiclassical expressions.

We remark that the level broadening used in Eq.\ (\ref{eq:dos})
ensures that the periodic orbit sums are truncated in such a way that
only orbits with periods $T_p$ shorter than the Heisenberg time
$t_{\rm H} = 2\pi\hbar/\Delta$ contribute.  
We note that one could
alternatively use a Lorentzian broadening \cite{ber94}.  For numerical
calculations, Gaussian broadened densities are much more convenient,
since one invariably deals with finite stretches of spectra, and
boundary effects are less pronounced due to faster decaying tails in
the Gaussian case.  

\section{The 2D Anderson model of localization}
\label{sec:am}

We performed numerical simulations within the 2D Anderson model of
localization\cite{and58}, by diagonalizing the Hamiltonian with the
help of the Lanczos algorithm\cite{C1-CW85}. In the site-basis the
model Hamiltonian with periodic boundary conditions is
\begin{equation}
  \widehat H= \sum_n|n\rangle\epsilon_n\langle n| + \sum_{n\neq
    m}|n\rangle t_{nm}\langle m|\,,
\end{equation}
where $|n\rangle$ represent the Wannier states at sites $n$ in the
$N\times N$ lattice. The on-site potential energies $\epsilon_n$ are
taken to be uniformly distributed between $-W/2$ and $+W/2$. The
hopping parameters $t_{nm}$ are non-zero only for nearest-neighbor
sites $n,m$ and we set the energy scale by choosing $t=1$ for these
sites. For convenience, we assume that the 2D model is embedded in 3D
and defines the $xy$-plane.

In the presence of a magnetic field the hopping parameters acquire an
additional factor $\exp i2\pi\phi/(\phi_0 {N})$, where $\phi$ is the
magnetic flux, which a periodic orbit encircles in the hopping
direction. This phase represents the Aharonov-Bohm effect on the
system with periodic boundary conditions under the magnetic flux.  We
use two magnetic fluxes, $\phi_x$ and $\phi_y$, corresponding to $x$-
and $y$-directions. The corresponding phase of the hopping parameter
is $\exp i2\pi[\phi_x/(\phi_0 N)+\phi_y/(\phi_0 N)]$.  For
completeness, we also study the influence of a homogeneous magnetic
field $B$ in $z$-direction. In this case the hopping parameters are
multiplied by $\exp \mp i2\pi B r_y/\phi_0$, when, e.g., hopping in
$x$-direction. $r_y$ is the $y$-coordinate of the site, and the sign
is different for opposite hopping directions.  To maintain the
appropriate periodicity of the boundary conditions, $B/\phi_0$ must
then be chosen as an integer multiple of $1/N$.  The hopping
parameters in $y$-directions do not change due to $B$, when we choose
the vector potential ${\bf A}$ in the Landau gauge ${\mathbf
  A}=(0,Bx,0)$.

The energy spectrum for a single realization of disorder still has an
energy dependent density of states. In order to study the universal
fluctuations, we thus need to ``unfold'' the spectrum \cite{hof93},
such that the original set of eigenvalues $\{E_\alpha\}$ is mapped to
a new set $\{\varepsilon_\alpha\}$, where
\begin{equation}
  \varepsilon_\alpha= \langle {\cal N}(E_\alpha,\phi) \rangle = 
{\cal N}(E_\alpha,\phi)-\widetilde{\cal N}(E_\alpha,\phi),
\end{equation}
where ${\cal N}(E,\phi) = \int_{-\infty}^{E} dE' d(E',\phi)$ is the
integrated density of states, and $\widetilde{\cal N}(E_\alpha,\phi)$ is
the fluctuating part of ${\cal N}(E,\phi)$.
In practise we computed $\langle {\cal N}(E_\alpha,\phi) \rangle$ by
fitting the ${\cal N}(E_\alpha,\phi)$ data to a second order
polynomial.  Then we set the value of the polynomial at $E_\alpha$ to
$\langle {\cal N}(E_\alpha,\phi) \rangle$.  This procedure works
particularly well for a relatively small number of eigenvalues, where
the mean level spacing $\Delta$ is almost a constant.  After
unfolding, we have $\Delta=1$.
We remark that an unfolding
based on a cubic spline interpolation \cite{hof93} does not work so
well in the present case.

The semiclassical approach applies to weakly disordered systems and
for parts of the spectrum, where the electron states spread throughout
the system. Thus the conductance $g= t_H/t_D$, with $t_D= L^2/\pi {\cal D}$
the Thouless time, should obey $g \gg 1$. However, in the infinitely
large 2D Anderson model, it is well-known that all states are
localized for any finite amount of disorder \cite{kra93,sch91}.
Nevertheless, for suitably weak disorder and at small systems,
one can find large regions in the spectrum for which $g \gg 1$
\cite{mul97}, such that we need not go to higher dimensions to test
the semiclassical results.  With zero flux the (unfolded) spectral
fluctuations of the 2D Anderson model in this limit of weak disorder
are described by the GOE of RMT \cite{hof93,zha97,dup91}.  Upon
increasing the flux there is a transition to GUE \cite{dup91}. In
order to test that we indeed are investigating a part of the spectrum
in which universality holds, we calculate the nearest-neighbor energy
level spacing distribution and check that the statistics for zero flux
is given by the Wigner-Dyson result for GOE, whereas for finite flux
or magnetic field we have the GUE result \cite{wig51}. In the
following sections we consider the dependence of the spectral
statistics on the magnetic flux $\phi_x\equiv\phi$ in the
$x$-direction.  The magnetic flux $\phi_y$ in the $y$-direction and
the homogenous magnetic field in $z$-direction are used as convenient
switches between GOE ($\phi_y=0$ and $B=0$) and GUE ($\phi_y\neq 0$ or
$B \neq 0$) behavior. We note that for weak magnetic flux
($\phi,\phi_y \ll \phi_0/4$), time-reversal symmetry is only weakly
broken and the statistical properties of the spectrum are described by
the Pandy-Mehta ensemble\cite{pan83,dup91}.

\section{The GOE to GUE transition}
\label{sec:goegue}

In this section, we will study the correlation functions of the
density of states $C_d(\phi)$, the density of level velocities
$C_v(\phi)$, and the density of matrix element correlations
$C_m(\phi)$ as functions of the external magnetic flux $\phi= \phi_x$.
Hence, we also have $\phi_y=0$, $B=0$. We shall always first consider
the semiclassical derivation of these correlations and then turn our
attention to a numerical computation within the 2D Anderson model.

\subsection{Density of states} 
\label{sec:goegue:dos}

We first consider correlations of the density of states, as defined in
Eq.\ (\ref{eq:dos}), and calculate the statistic
\begin{equation}
  C_{\rm d}(\phi) = \Big\langle \Big| \widetilde d(E,\phi)
  \Big|^2\Big\rangle_{E}\,,
\end{equation}
where $\langle \cdots\rangle_E$ denotes an average over a suitably
chosen energy interval as explained in the last section.  
Within the diagonal approximation~\cite{ber94} we obtain
\begin{eqnarray}
  \nonumber C_{\rm d}(\phi) = \frac{1}{2\pi^2\epsilon^2}
  \sum_{\nu=-\infty}^\infty \Big \{ 1& -& \frac{\sqrt{\pi}}{2}
  \left(\frac{\nu}{\delta}\right)^2 \exp\Big(\frac{\nu}{\delta}\Big)^4
  \, \mbox{erfc}\Big(\frac{\nu}{\delta}\Big)^2\\ &-&
  \frac{\sqrt{\pi}}{2} z \exp(z^2) \,\mbox{erfc}(z) \Big\}
\label{eq:cd}
\end{eqnarray}
with $z = (\nu-2\phi/\phi_0)^2/\delta^2$, 
$\delta^2 = \epsilon/\pi^2\lambda\hbar$
and $\hbox{erfc}(z)$ the complementary error function\cite{gra94}.
This expression describes the crossover of the spectral
properties from GOE to GUE behavior, as the flux $\phi$ is varied.
A corresponding expression for a transition driven
by a magnetic field was given in \cite{boh95}.
Note that Eq. (\ref{eq:cd}) is periodic in $\phi$
with period $\phi_0/2$.  Eq.\ (\ref{eq:cd}) can be further simplified
in the limit of small $\delta$ (with  $\epsilon > 1$).
We consider two cases, namely $\phi = 0$ and $\phi = \phi_0/4$.  In
the first case, the system exhibits fluctuations described by the GOE, 
in the second case the fluctuations are described by
the GUE.  We then have
\begin{equation}
C(\phi)
\simeq
\frac{2}{\beta}\,
\frac{1}{4\pi^2\epsilon^2} \,,
\end{equation}
where $\beta = 1$ in the GOE and $\beta = 2$ in the GUE.
It must be emphasized that one requires
$\epsilon \gtrsim 1$ for Eq.\ (\ref{eq:cd}) to hold.
This ensures that only orbits with periods
$T_p < t_{\rm H}$ contribute to (\ref{eq:gutz}).
For small values of the level broadening,
the diagonal approximation used in deriving
(\ref{eq:cd}) ceases to be valid~\cite{bog96}.
On the other hand,
in the limit of $\epsilon  \ll 1$, one has \cite{kea91}
\begin{eqnarray}
C(\phi)
&\simeq&  \Big\langle\sum_\alpha \delta_\epsilon^2[E-E_\alpha(\phi)]
\Big\rangle
\simeq  \frac{1}{2\sqrt{\pi}\epsilon}\,,
\label{eq:cd-lim}
\label{eq:smalleps}
\end{eqnarray}
which is independent of $\phi$.
In summary, one obtains for GOE and GUE
\begin{equation}\label{eq:cd:cross}
C(\phi) =
\left\{
\begin{array}{ll}
\frac{\displaystyle 1}{\displaystyle 2\sqrt{\pi}\epsilon} &
\mbox{for $\epsilon < \epsilon_{\rm c}$},\\
\frac{\displaystyle 2}{\displaystyle \beta}
\frac{\displaystyle 1}{\displaystyle 4\pi^2\epsilon^2} &
\mbox{for $\epsilon > \epsilon_{\rm c}$}\,.
\end{array}
\right .
\end{equation}
Thus the crossover between these two limiting behaviors occurs
at $\epsilon_{\rm c} \simeq \pi^{-3/2}/\beta$.

\subsubsection*{Numerical results for the density of states}
\label{sec:goegue:dos:num}

We obtained numerical data from the 2D Anderson model for 90 samples
of different realizations of disorder with $W=2.4$, using flux values
$\phi/\phi_0= 0, 0.007, 0.014, \ldots, 0.497$.  Larger values are not
needed because of the periodicity of $C_d$ in $\phi_0/2$. There were
$27 \times 27$ sites in the system. For each disorder realization we
computed $100$ subsequent energy eigenvalues $E_i \in [-3.4, -1.9]$,
thereby avoiding contributions from localized states in the band tails
and from nearly ballistic states at the band center.  We remark that
the mean density of levels is already nearly constant for this
interval and thus the second order polynomial is ideal for the
unfolding procedure.  After unfolding these eigenvalues, we calculated
the Wigner-Dyson statistics $P(s)$ for nearest-neighbor level
spacings. As shown in Fig.\ \ref{fig:enhist:W0240}, we find for flux
$\phi=0$ that $P(s)$ follows the GOE behavior. For flux values close
to $\phi_0/4$, we have $P(s)$ of the GUE.  Thus with this choice of
parameters we are indeed in the ergodic regime of the model as
required.

The comparison between the results for the Anderson model, averaged
over all disorder realizations, and the semiclassical approximation
with different broadening values $\epsilon$ in units of $\Delta$ is
shown in Fig.~\ref{fig:cd}. The agreement is the best for $\epsilon
\gtrsim 1$, as expected. For smaller values there are deviations near
the GOE cases $\phi=0$ and $\phi=0.5 \phi_0$.  
The constant $\lambda=1.21$ used in plotting Fig.~\ref{fig:cd}
was determined from the statistics of level velocities, as we explain
below in section \ref{sec:goegue:dol}. We emphasize that in Fig.\ 
\ref{fig:cd} and throughout the rest of this paper, we have not
symmetrized our data with respect to the periodicity in $\phi_0$. Thus
the slight deviations from periodicity at $\phi_0/2$ reflect the
accuracy of our data.

In Fig.\ \ref{fig:cd:cross} we show the small $\epsilon$-behavior of
$C_d$.  The crossover, predicted in Eq.\ (\ref{eq:cd:cross}) at
$\epsilon_{\rm c} \simeq \pi^{-3/2}/\beta\approx 0.18/\beta$, can be seen
to occur between the values $0.03 <\epsilon< 0.8$ for the GOE, and
$0.03 <\epsilon< 0.15$ for the GUE. The upper limits of the intervals
in each case can be considered as lower boundaries for the validity
range of the diagonal approximation. The upper validity range of the
diagonal approximation can also be inferred from Fig.\ 
\ref{fig:cd:cross} to be close to $4.5$ for GOE and $1.7$ for GUE.

\subsection{Density of level velocities} 
\label{sec:goegue:dol}

Next we consider fluctuations of the density of level velocities, 
and compute the statistic
\begin{equation}
  C_{\rm v}(\phi) = \Big\langle \Big| \widetilde d_{\rm v}(E,\phi)
  \Big|^2\Big\rangle_{E}\,.
\end{equation}
Within the diagonal approximation, we obtain
\begin{eqnarray}
\label{eq:cv}
C_{\rm v}(\phi)&&= \frac{\lambda\hbar}{\epsilon}
\sum_{\nu=-\infty}^\infty\nonumber \Big\{
\Big[1+4\Big(\frac{\nu}{\delta}\Big)^4\Big]\sqrt{\pi}
\exp\Big(\frac{\nu}{\delta}\Big)^4 \,
\mbox{erfc}\Big(\frac{\nu}{\delta}\Big)^2\nonumber\\
&&-4\Big(\frac{\nu}{\delta}\Big)^2 -\big[1+4z^2]\sqrt{\pi}\exp(z^2)
\,\mbox{erfc}(z) -4z\Big\}
\end{eqnarray}
with $z$ and $\delta$ as in Eq.\ (\ref{eq:cd}). For small $\delta$
(and with $\epsilon > 1$),
one obtains the following limiting behaviour
\begin{equation}
  C_{\rm v}(\phi) \simeq \left\{
\begin{array}{cl}
  0 & \mbox{for $\phi = 0$\,,}\\ {\sqrt{\pi} \lambda\hbar}/{\epsilon}
  &\mbox{for $\phi = \phi_0/4$\,.}
\end{array}
\right .
\label{eq:cv-lim}
\end{equation}
Alternatively, in the limit of very small $\epsilon$, we 
obtain in analogy with Eq. (\ref{eq:smalleps})
\begin{eqnarray}
  C_{\rm v}(\phi) 
&\simeq& \Big\langle\sum_\alpha \Big( \frac{\partial E_\alpha
    }{\partial \phi}\Big)^2\, \delta_\epsilon^2[E-E_\alpha(\phi)]
  \Big\rangle
\simeq \frac{\mu_{\rm
      diag}^2}{2\sqrt{\pi}\epsilon}\,,
\end{eqnarray}
where $\mu_{\rm diag}^2$ is the variance of the level velocities
$\partial E_{\alpha} / \partial \phi$.  For $\beta = 2$, we
have \cite{wilk90}
\begin{equation}\label{eq:cv:off1}
  \mu^2_{\rm diag}(E) = \mu^2_{\rm off}(E)\,,
\end{equation}
where
\begin{equation}
  \mu^2_{\rm off}(E) = \left\langle \left(\frac{\partial
        H}{\partial\phi} \right)_{\alpha{\alpha'}}^2 \right\rangle
  \rule[-5mm]{0.075mm}{11mm}_{ \stackrel{\;\scriptstyle E_\alpha
      \simeq E_{\alpha'} \simeq E} {\;\scriptstyle\hspace{-0.8cm}
      \alpha \neq {\alpha'}}}\,.
\label{eq:off:def}
\end{equation}
With $\mu^2_{\rm off}(E) = 2\pi\hbar \lambda$
(see section V)
we obtain for the GUE case ($\beta = 2$)
\begin{equation}\label{eq:cv:small}
  C_{\rm v}(\phi) = \sqrt{\pi}\, \frac{\lambda \hbar}{\epsilon}\,.
\end{equation}
This implies that the semiclassical result of Eq.\ (\ref{eq:cv-lim}),
obtained within the diagonal approximation, remains valid for small
$\epsilon$ [as opposed to the estimate (\ref{eq:cd-lim})].
We remark that while this
is true for the GOE ($\phi=0, \phi_0/2$) and GUE ($\phi=\phi_0/4$)
cases, it is no longer true in the transition regime
\cite{dup91}. It will be seen in the next section
that similar arguments apply to fluctuations
of matrix elements.

\subsubsection*{Numerical results for the density of level velocities}
\label{sec:goegue:dol:num}

Using the same data as for the density of states correlations, we
computed $C_v$ for the Anderson model with different broadenings, as
shown in Fig.~\ref{fig:cv}. In this case the agreement with the
semiclassical approximation is good even around $\phi=0$ and
$\phi=\phi_0/2$, i.e.\ in the GOE case. 
We remark that the shoulders
visible in Fig.\ \ref{fig:cv} around $\phi= 0.07\phi_0$ and
$0.43\phi_0$ for the semiclassical expressions at small $\epsilon$ are
an artefact of our approximation for $\epsilon < \Delta$. 

The parameter $\lambda$ was determined from the small
$\epsilon$-behavior of the $C_v$ by fitting the numerical results to
Eq.\ (\ref{eq:cv:small}) as shown in Fig.~\ref{fig:cv:small}.  The
agreement of the small $\epsilon$-behavior of the numerical data with
Eq.\ (\ref{eq:cv:small}) is rather good. Indeed, the agreement is good
for all values of $\epsilon$, as expected from Eq.\ (\ref{eq:cv-lim})
and discussed above.
An alternative way is to compute a histogram for the level velocities
in the unitary case and to use 
Eq.\ (\ref{eq:cv:off1})
and the estimate $\mu^2_{\rm off}(E) = 2\pi\hbar \lambda$.
This is shown in Fig.~\ref{fig:velhist1}. Both
methods do not give exactly the same value of $\lambda$ due to
numerical accuracy and the limited number of samples.  With the former
method we estimate a value $\lambda=1.2\pm0.1$, and with the latter one
$\lambda=1.4\pm0.2$, where the error limits represent the standard
deviation of the values obtained for different realizations of
disorder.  We have chosen the value $\lambda=1.21$ such that the overall
agreement of each correlation function in Fig.~\ref{fig:cv} is as good
as possible for all $\phi$ and all $\epsilon \gtrsim 0.3$. We
emphasize that such an agreement is very sensitive on the actual value
of $\lambda$ chosen.  Furthermore, we need to assume that $\lambda$
remains constant for all $\phi$. As we will show later, this
assumption is at least questionable for the Anderson model.

\subsection{Density of matrix elements}
\label{sec:goegue:dom}

In this section we turn to fluctuations of expectation values
and consider the statistic
\begin{equation}
  C_{\rm m}(\phi) = \Big\langle \Big| \widetilde d_{\rm m}(E,\phi)
  \Big|^2\Big\rangle_{E}
\end{equation}
and obtain, again in the diagonal approximation,
\begin{eqnarray}
  C_{\rm m}(\phi) &=& \frac{\sigma_{\rm off}^2(E)} {2\sqrt{\pi}\epsilon}
  \sum_{\nu=-\infty}^\infty\nonumber\\ 
  &\times&\Big\{\exp\Big(\frac{\nu}{\delta}\Big)^4\,
  \mbox{erfc}\Big(\frac{\nu}{\delta}\Big)^2 +
  \exp(z^2)\,\mbox{erfc}(z)\Big\}
\label{eq:cm}
\end{eqnarray}
with $z$ and $\delta$ as in Eq.\ (\ref{eq:cd}). 
Moreover, $\sigma_{\rm off}^2(E)$ is the
variance of  non-diagonal matrix elements
\begin{equation}\label{eq:cm:var}
  \sigma^2_{\rm off}(E) = \left\langle |A_{\alpha{\alpha'}}|^2
 \right\rangle \rule[-7mm]{0.075truemm}{10mm}_{ \stackrel{\;\scriptstyle
     E_{\alpha\phantom{'}} \simeq E_{\alpha'} \simeq E}
{\;\scriptstyle\hspace{
-0.8cm}
      \alpha \neq {\alpha'}}}\,.
\end{equation}
Correspondingly, $\sigma_{\rm diag}^2(E,\phi)$ is the variance
of diagonal matrix elements. Unlike $\sigma_{\rm off}^2$
it depends on the value of the flux $\phi$.
In the limiting cases of GOE and GUE,
the variances are related as
\begin{equation}
 \label{eq:2}
 \sigma^2_{\rm diag}(E,\phi) = \frac{2}{\beta}\sigma^2_{\rm off}(E)\,.
 \end{equation}
In the limit of small
$\delta$, one obtains for GOE and GUE,
\begin{equation}
  C_{\rm m}(\phi) \simeq \frac{2}{\beta}\frac{\sigma_{\rm off}^2}
  {2\sqrt{\pi}\epsilon} \,.
\label{eq:cm-lim}
\end{equation}
We shall now argue that these results, derived assuming
$\epsilon\gtrsim 1$, remain valid in the limit of small $\epsilon$.
Proceeding as in the previous section, we obtain for small $\epsilon$
\begin{equation}\label{eq:cm:small}
  C_{\rm m}(\phi) \simeq \frac{\sigma_{\rm
      diag}^2}{2\sqrt{\pi}\epsilon} = \frac{2}{\beta}\,
  \frac{\sigma_{\rm off}^2}{2\sqrt{\pi}\epsilon}\,,
\end{equation}
which is the same as Eq.\ (\ref{eq:cm-lim}) calculated for $\epsilon
\gtrsim 1$.

\subsubsection*{Numerical results for the density of matrix elements}
\label{sec:goegue:dom:num}

We computed eigenvalues and the expectation values of the diagonal
matrix elements $x_{nn}$ for the dipole moment operator $\hat{x}$ in
the site-basis for $69$ different realizations of disorder $W=2.4$ in
the Anderson model at flux $\phi/\phi_0= 0, 0.007, \ldots, 0.497$. We
obtained $C_m$ with different broadenings $\epsilon$ as shown in
Fig.~\ref{fig:cm}. Here the agreement is reasonable, but not as good
as in the two previous cases. We note that the small $\phi$ behavior
is much better described by the universal $\nu=0$ term than by the
complete expression of Eq.\ (\ref{eq:cm}). 

We emphasize that for the present correlation, we had to determine
{\em two} constants describing the system, namely, $\lambda$ and
$\sigma^2_{\mathrm{off}}$. This makes it even more important to have
various independent ways of computing them.  The variance
$\sigma^2_{\mathrm{off}}$ of the off-diagonal matrix elements can be
determined from the diagonal elements in a similar way as the
determination of the diffusion constant from the level velocities.
Namely, we can use the small $\epsilon$-behavior of Eq.\ 
(\ref{eq:cm-lim}) as shown in Fig.~\ref{fig:cm:small}. Interestingly,
we find that although Eq.\ (\ref{eq:cm-lim}) is expected to remain
valid for $\epsilon\simeq 1$, there are already strong deviations of
our numerical data from the behavior predicted by Eq.\ 
(\ref{eq:cm-lim}). This may indicate that the approximations used in
the derivation of Eq.\ (\ref{eq:cm}) are less reliable for the matrix
element correlations than for density of states and velocity
correlations.
We can also use the histogram of the diagonal matrix elements as shown
in Fig.\ \ref{fig:mehist:goegue}.  Both methods give slightly
different values for $\sigma^2_{\mathrm{off}}$ in GOE and GUE, whereas
we assumed in the derivation of the semiclassical formulae that
$\sigma^2_{\mathrm{off}}$ is independent on the magnetic flux. For GOE
we obtain a value around $\sigma^2_{\mathrm{off}}=0.7\pm0.1$ and for
GUE $\sigma^2_{\mathrm{off}}=0.8\pm0.2$.  Both estimates are
compatible within the error limits, though.  In
Fig.~\ref{fig:cm:small}, we choose $\sigma^2_{\mathrm{off}}=0.65$ in
order to get the best overall agreement between Eq.\ (\ref{eq:cm}) and
our numerical results.  Also, we have again used $\lambda=1.21$ as an
estimate of $2{\cal D}/L^2$ as in the previous sections.

Keeping in mind the sensitivity of the expressions (\ref{eq:cd}),
(\ref{eq:cv}), and (\ref{eq:cm}) to the actual values of $\lambda$ and
$\sigma^2_{\mathrm{off}}$, we can conclude this section by noting that
our numerical data for the 2D Anderson model in the ergodic regime
show the main features predicted for the correlations and
convincingly exhibit the GOE to GUE transition.

\section{Parametric statistics}
\label{sec:para}

In this section, we will study the parametric correlation functions of
the density of states $K_d(\Delta\phi)$, the density of level
velocities $K_v(\Delta\phi)$, and the density of matrix elements
$K_m(\Delta\phi)$ as functions of the difference in external magnetic
flux $\Delta\phi= \Delta\phi_x$, averaged over different flux values
$\phi$.  Since, as studied in the previous section, the spectral
properties change from GOE to GUE as $\phi$ is varied, we introduce an
additional flux $\phi_y= \phi_0/4$ in the transverse direction, so as
to have spectral statistics according to the GUE for all values of
$\phi$. Again, we shall start by first considering the semiclassical
derivation of these parametric correlations and afterwards compare to
numerical data from the 2D Anderson model.

\subsection{Density of states}
\label{sec:para:dos}

For the parametric case we define
\cite{sim93a}
\begin{equation}
  K_{\rm d}(\Delta\phi) = \langle \widetilde d(E,\phi) \widetilde
  d^\ast(E,\phi+\Delta\phi) \rangle_{E,\phi}\,,
\end{equation}
where $\langle \cdots \rangle_{E,\phi}$ denotes an average over $E$
and $\phi$.  One obtains within the diagonal approximation
\begin{eqnarray}
  K_{\rm d}(\Delta \phi) &=& \sum_{\nu=-\infty}^\infty
  \frac{1}{4\pi^2\epsilon^2} \left\{ 1-\sqrt{\pi} \,z
    \,\exp(z^2)\,\mbox{erfc}(z) \right\}\,
\label{eq:kd}
\end{eqnarray}
with $z = (\nu+\Delta\phi/\phi_0)^2/\delta^2$ and $\delta^2 = \epsilon
/\pi^2\hbar \lambda$.

\subsubsection*{Numerical results for the density of states}
\label{sec:para:dos:num}

We computed $69$ realizations of disorder for the 2D system with $27
\times 27$ sites and a disorder strength $W=1.7$, using the same part
of the spectrum as previously and flux values $\phi/\phi_0= 0, 0.01,
0.02, \ldots, 1.0$.  $P(s)$ reflects the GUE, as in Fig.\ 
\ref{fig:enhist:W0240}, for all values of $\phi$ due to the additional
transverse flux $\phi_y$.  In Fig.~\ref{fig:kd}, we show the
comparison between the semiclassical expression (\ref{eq:kd}) and the
numerical data. The agreement is very good for all values
of $\epsilon$.

The parameter $\lambda$ was determined in the same way as in the GOE to
GUE transition in section \ref{sec:goegue}. Because the system had
been made unitary by introducing an additional flux $\phi_y$, Eq.\ 
(\ref{eq:cv:small}) is valid for all values of $\phi$.  Consequently,
the fitting procedure for the small $\epsilon$-values should give the
same $\lambda$ for all the flux values, and the histogram of the level
velocities should have the same variance. However, we found
differences, which cannot be explained only by the error bars.  This
has been illustrated in Fig.~\ref{fig:diff:gue}.  The value
$\lambda=2.5$, used in Fig.\ \ref{fig:kd} was chosen such that the
agreement is the best for all $\Delta\phi$, all $\epsilon$ and all
three parametric correlations.

We also used $W=2.4$ as in section \ref{sec:goegue} for the GOE to GUE
transition and computed the parametric correlations. But in this case
the agreement between the semiclassical theory and the data obtained
from the Anderson model is slightly less convincing than with $W=1.7$.

\subsection{Density of level velocities}
\label{sec:para:dol}

For the parametric correlation of the density of level velocities, we
define \cite{sim93a}
\begin{equation}
  K_{\rm v}(\Delta\phi) = \langle \widetilde d_{\rm v}(E,\phi)
  \widetilde d^\ast_{\rm v}(E,\phi+\Delta\phi) \rangle_{E,\phi}\,.
\end{equation}
Within a semiclassical approach, we obtain
\begin{eqnarray}
  K_{\rm v}(\Delta \phi) &=& \frac{\lambda\hbar}{\epsilon}\sum_{\nu =
    -\infty}^\infty\nonumber \\ &&\hspace{-1cm}
  \times\Big\{\big(1+4z^2\big) \sqrt{\pi} \exp(z^2) \,\mbox{erfc}(z)
  -4z\Big\}
\label{eq:kv}
\end{eqnarray}
with $z$ and $\delta$ as in Eq.\ (\ref{eq:kd}) and for Gaussian
broadening.  This expression is periodic in $\Delta\phi$ with
period $\phi_0$.  It has previously been derived in \cite{ber94},
using Lorentzian broadening, see also \cite{walk95}. Comparing
the $\nu=0$ term of Eq.\ (\ref{eq:kv}) with the corresponding
expression
\begin{equation}
  \frac{\mu^2_{\rm off}(E)}{2\pi\epsilon} \Big\{\big(1+4z^2\big)
  \sqrt{\pi} \exp(z^2) \,\mbox{erfc}(z) -4z\Big\}\,
\end{equation}
obtained from a Brownian motion model \cite{wilk96}, we have
$\mu^2_{\rm off}(E) = 2\pi\hbar \lambda$
(compare section IV).

\subsubsection*{Numerical results for the density of level velocities}
\label{sec:para:dol:num}

Using the same data as in section \ref{sec:para:dos:num} for the
density of states, we computed the parametric statistics for the
density of level velocities for the Anderson model. The comparison
with Eq.\ (\ref{eq:kv}) can be seen in Fig.~\ref{fig:kv}.  The
agreement with the semiclassical approximation is again very good.  We
remark that the overestimation of the minima in $K_{\rm v}$ around
$\Delta\phi= 0.1\phi_0$ and $0.9\phi_0$ for the semiclassical
expressions at small $\epsilon\lesssim 0.1$ is an artefact of the
diagonal approximation \cite{meh98c}.

\subsection{Density of matrix elements}
\label{sec:para:dom}

Lastly, we consider the parametric correlation $K_{\rm m}(\Delta
\phi)$ of matrix elements, i.e.,
\begin{equation}
  K_{\rm m}(\Delta \phi) = \langle \widetilde d_m(E,\phi) \widetilde
  d_m^\ast(E,\phi+\Delta\phi) \rangle_{E,\phi}\,.
\end{equation}
As before we obtain
~\cite{wilk96}
\begin{equation}
  K_{\rm m}(\Delta \phi) = \frac{\sigma^2_{\rm
      off}}{2\sqrt{\pi}\epsilon} \sum_{\nu=-\infty}^\infty \exp(z^2)\;
  \mbox{erfc}(z)
\label{eq:km}
\end{equation}
within the diagonal approximation and with $z$ and $\delta$ as in Eq.\ 
(\ref{eq:kd}). 
We have assumed that the mean density
of states $\langle d\rangle$ is essentially energy- and
flux-independent.  Moreover, we have neglected the
energy-dependence of the off-diagonal variance.

\subsubsection*{Numerical results for the density of matrix elements}
\label{sec:para:dom:num}

In Fig.~\ref{fig:km}, we show the comparison between semiclassical and
numerical results for the parametric statistics of the matrix elements
of the dipole moment operator using the same data as for the two
previous parametric correlations.  The agreement here is even better
than in the GOE to GUE transition. This is noteworthy, because of the
large discrepancies between the values of $\lambda$ for different flux
values (cp.\ Fig.~\ref{fig:diff:gue}) which we neglected in the
semiclassical derivation of Eq.\ (\ref{eq:km}). The off-diagonal
variance $\sigma_{\mathrm{off}}^2$ was determined in the same way as
in section \ref{sec:goegue} for the GOE to GUE transition, giving
$\sigma_{\mathrm off}^2=0.50\pm0.05$.  We get different values for
different flux values as for the diffusion constant, but the
variations are much smaller. By calculating directly the variance of
the matrix elements between nearest-neighbor sites we get a slightly
larger value $\sigma_{\mathrm off}^2=0.65\pm0.05$. Here, the error
bars represent the deviations from the average value for different
flux values.

Thus in summary, we find that as in section \ref{sec:goegue}, the
general behavior of the data obtained for the 2D Anderson model in the
GUE case is very well reproduced by the semiclassical expressions
(\ref{eq:kd}), (\ref{eq:kv}), and (\ref{eq:km}). In fact, the
agreement is even better than in section \ref{sec:goegue}.

\section{Distributions}
\label{sec:histo}

The distributions of level velocities \cite{sim93a}, shown in
Fig.~\ref{fig:velhist1}, and of the diagonal matrix elements of the
dipole moment operator, in Fig.~\ref{fig:mehist:goegue}, are well
approximated by Gaussian distributions, as predicted in random matrix
theory.  According to Eq.\ (\ref{eq:cm:var}) the variance of the
matrix elements in the GOE case ($\phi=0$) should be approximately two
times larger than in the GUE case ($\phi\approx\phi_0/4$). We obtain a
factor of $\sigma^2_{\mathrm diag}(\phi=0)/\sigma^2_{\mathrm
  diag}(\phi\approx\phi_0/4) \approx
(1.3\pm0.2)/(0.8\pm0.1)=1.6\pm0.5$ in agreement with this prediction,
although the standard deviations are quite large.  We again emphasize
that the level spacing distributions obey the Wigner-Dyson statistics,
predicted in random matrix theory, as shown in
Fig.~\ref{fig:enhist:W0240} for all the disorders and magnetic fields
chosen in our work.

We also calculated the distributions of the off-diagonal elements
$A_{\alpha{\alpha'}}$ with $E_\alpha \simeq E_{\alpha'}$. 
We find that their distribution is
also well approximated by a Gaussian as shown in Fig.\ 
\ref{fig:off-diag}.  The corresponding variance
$\sigma^2_{\mathrm{off}}$ should be independent of the magnetic flux.
This is approximately true for our data. With disorder $W=2.4$ we get
$\sigma^2_{\mathrm{off}}=0.8\pm0.2$ in GOE ($\phi_x=\phi_y=0$), and
$0.9\pm0.2$ in GUE ($\phi_x=\phi_0/4$, $\phi_y=0$) and with $W=1.7$,
we find $\sigma^2_{\mathrm{off}}=0.7\pm0.2$ at $\phi_x=0$,
$\phi_y=\phi_0/4$ and $0.6\pm 0.2$ at $\phi_x=\phi_y=\phi_0/4$. The
error bars represent again the standard deviations of the values
obtained for different realizations of disorder.

\section{Conclusions}
\label{sec:concl}

In this paper, we have reported on extensive calculations
of smoothed correlation functions in the 2D
Anderson model of localization.   
We have  calculated
correlation functions of energy levels, their
parametric derivatives and of diagonal matrix elements
in the metallic regime  ($g\gg 1$).
For two cases, namely for parametric correlations
and for fluctuations in the transition
regime between GOE and GUE, we have
presented detailed comparisons of
our numerical results with semiclassical
theory, focussing on the dependence
of the fluctuations on the
level broadening.

Our results can be summarized as follows.
First, one expects the semiclassical
theory to be appropriate for
level-broadenings  
in the range of
$1 < \epsilon \ll g$
(with $\epsilon$ in units of $\Delta$).
Comparison with asymptotic expressions
for small $\epsilon$ 
[Eqs. (\ref{eq:smalleps}), (\ref{eq:cv:small})
and (\ref{eq:cm:small})] shows
that
the lower bound actually extends to $\epsilon_{\rm c} \simeq
\pi^{-3/2}/\beta$ for density-of-states
fluctuations. In the case of fluctuations
of level velocities and matrix elements, moreover,
the diagonal approximation remains valid
for arbitrarily small $\epsilon$.
This is simply due to the fact that the
additional factors in Eqs. (\ref{eq:vdos})
and (\ref{eq:mdos}) are essentially random and help
to suppress off-diagonal
contributions to (\ref{eq:diag}). 
Our numerical results verify these conclusions.

Second, at large values of $\epsilon$ we
observe deviations from the universal
theoretical results, as expected.
This is evident in Figs.
\ref{fig:cd:cross}, \ref{fig:cv:small}
and \ref{fig:cm:small}. The value
of the conductance in this case
is $g = 12 \pm 3$. Interestingly,
in Fig. \ref{fig:cm:small} in 
particular,
we observe deviations from
the universal prediction
at considerably smaller values
of $\epsilon$. From this
we conclude
that fluctuations of
matrix elements are particularly
sensitive to non-universal
effects. This is consistent
with the following observation.
In the universal regime,
the semiclassical expressions
derived in this paper
should be dominated 
by those terms 
for which $|\nu -2\phi/\phi_0|$
is minimal. However, in
the case of matrix element
fluctuations, non-universal
contributions are
particularly large
(compare Fig. \ref{fig:cm}).
This is not surprising
since it can be shown that
short periodic orbits
make large, non-universal
contributions to $C_m(\phi)$.

Third, in the case of parametric
fluctuations 
(Figs. \ref{fig:kd}, \ref{fig:kv}
and \ref{fig:km}) we observe excellent
agreement with the semiclassical
predictions. This is due
to the fact that (i) these
numerical results are averaged
over a considerably larger
ensemble and (ii) that
the conductance is larger
($g = 24 \pm 5$).

Fourth, we emphasize that in our
case the parameters $g$ and
$\sigma_{\rm off}^2(E)$
are found to depend on the
magnetic flux 
(compare Fig. \ref{fig:diff:gue}).
The flux dependence turned out to be more prominent with the smaller disorder 
strength we used. That is why our numerical results for the correlations in 
the GOE to GUE transition 
agrees better with the semiclassical formulae with $W=2.4$ than 
with $W=1.7$, even if the conductance is smaller in the former case. 
Within the framework of the semiclassical
theory $g$ and $\sigma_{\rm off}^2(E)$ are expected
to be independent of $\phi$
since an Aharonov-Bohm flux
does not change the classical mechanics.

Fifth, we have verified the
relation between
the variances of diagonal
and non-diagonal matrix elements
in the GOE and GUE.
The agreement of our
numerical results with the prediction
is reasonably good\cite{note}.

In summary, we  have shown
to which extent fluctuations
in the 2D
Anderson model
are accurately described
by universal semiclassical 
formulae. We have found,
in particular,
that the fluctuations
depend sensitively
on the level-broadening
and that this dependence
can be used to
assess consistency with
RMT, as originally suggested in
\cite{wilk88}.
This is particularly important
for the following reason.
In order to test
recent predictions
\cite{falko} on the
effect of incipient
localization on
the fluctuations of
wave-function amplitudes
in the 2D
Anderson model
it is essential to
have an accurate
and {\em quantitative}
understanding  of
the metallic regime.

\acknowledgements V.U.\ would like to thank F.\ Milde for 
help with using the Lanczos algorithm
and thankfully acknowledges
financial support by the DAAD. V.U. and R.A.R. gratefully acknowledge
support by the DFG as part of Sonderforschungsbereich 393.


\newpage \newcommand{\figwidth}{0.9\columnwidth}


\begin{figure}

  \centerline{\psfig{file=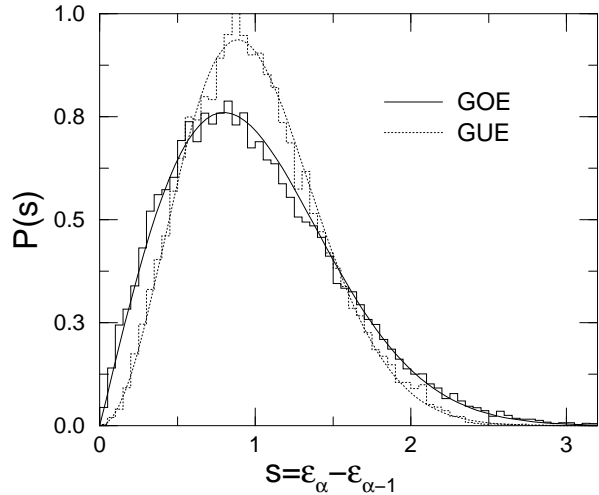,width=\figwidth}}
\caption{\label{fig:enhist:W0240}
  Histograms for energy level spacings of the unfolded energies for
  all samples with disorder $W=2.4$ and a system size $N^2 = 27^2$.
  The (smooth) lines denote the GOE (solid) and GUE (dotted)
  Wigner-Dyson distributions \protect\cite{wig51} for $\phi=0$ and
  $\phi/\phi_0= 0.25 \pm 0.05$, respectively.}
\end{figure}


\begin{figure}

  \centerline{\psfig{file=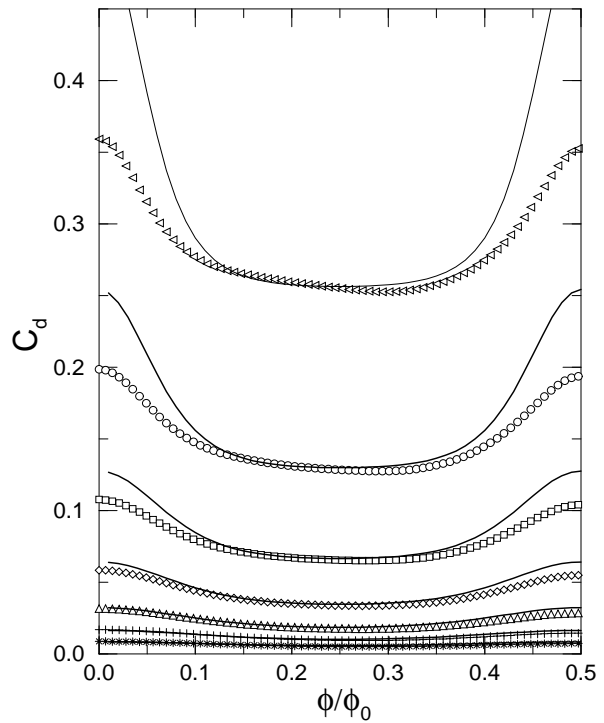,width=\figwidth,angle=0}}
\caption{\label{fig:cd}
  GOE to GUE transition for density of states correlations according
  to Eq.\ (\protect\ref{eq:cd}) (solid lines) and corresponding
  results from the numerical simulations of the Anderson model
  (symbols). The parameter $\lambda=1.21$ and $\epsilon =
  0.316$ ($\lhd$), $0.447$ ($\circ$), $0.631$ ($\Box$), $0.891$
  ($\Diamond$), $1.26$ ($\bigtriangleup$), $1.78$ ($+$), $2.51$
  ($\ast$). }
\end{figure}


\begin{figure}

  \centerline{\psfig{file=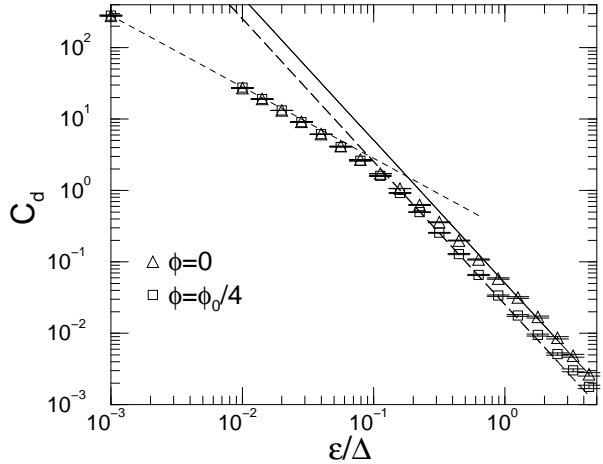,width=\figwidth,angle=0}}
\caption{\label{fig:cd:cross}
  Small $\epsilon$-behavior of $C_d$ with $W=2.4$.  The solid line
  indicates $1/(2\pi^2\epsilon^2)$, the long dashed line is
  $1/(4\pi^2\epsilon^2)$ and the short dashed line denotes
  $1/(2\protect\sqrt{\pi}\epsilon)$.}
\end{figure}


\begin{figure}

  \centerline{\psfig{file=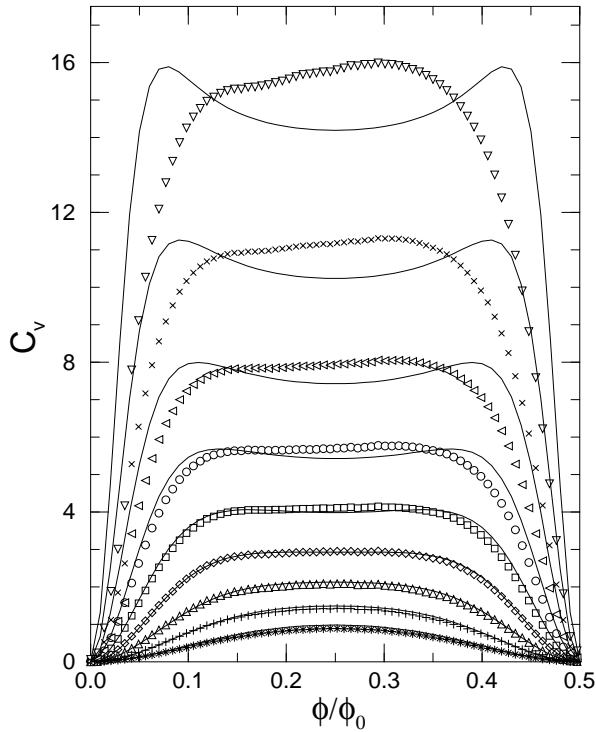,width=\figwidth,angle=0}}
\caption{\label{fig:cv}
  GOE to GUE transition for level velocity correlations according to
  Eq.\ (\protect\ref{eq:cv}) (solid lines) and corresponding results
  from the numerical simulations of the Anderson model (symbols). The
  parameters are the same as in Fig.~\protect\ref{fig:cd}. We
  additionally include the broadenings $\epsilon= 0.158$
  ($\bigtriangledown$) and $0.224$ ($\times$).}
\end{figure}


\begin{figure}
\centerline{%
  \psfig{file=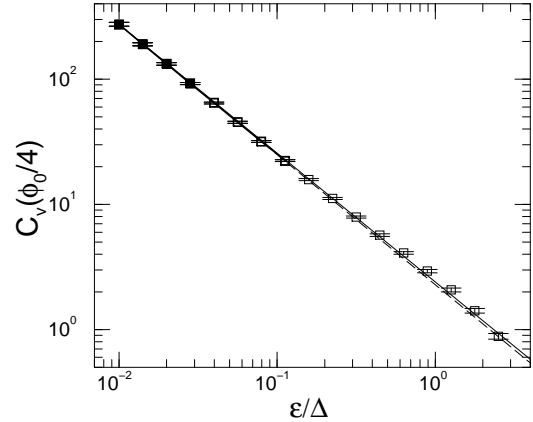,width=\figwidth,angle=0}}
\caption{\label{fig:cv:small}
  Determination of the parameter $\lambda$ from the small
  $\epsilon$-behavior of $C_v$ at $W=2.4$.  By fitting the four first
  points ($\blacksquare$) on the left to Eq.\ 
  (\protect\ref{eq:cv:small}) one gets a value $\lambda=1.22\pm0.02$,
  whereas fitting the first eight points (bold $\square$) gives
  $\lambda=1.30\pm0.01$. The difference between a plot of Eq.\ 
  (\protect\ref{eq:cv:small}) with $\lambda=1.22$ (thin dashed line) and
  $\lambda=1.3$ (thin solid line) is very small.}
\end{figure}


\begin{figure}
  \centerline{\psfig{file=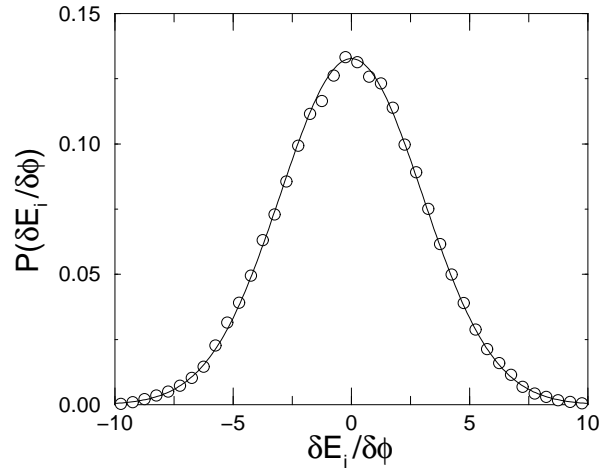,width=\figwidth}}
\caption{\label{fig:velhist1}
  Distribution of level velocities averaged over flux values
  $\phi/\phi_0=0.175, \ldots , 0.329$ and $90$ different realizations of
  disorder for $W=2.4$. The line represents a fit by a Gaussian
  distribution.  }
\end{figure}


\begin{figure}

  \centerline{\psfig{file=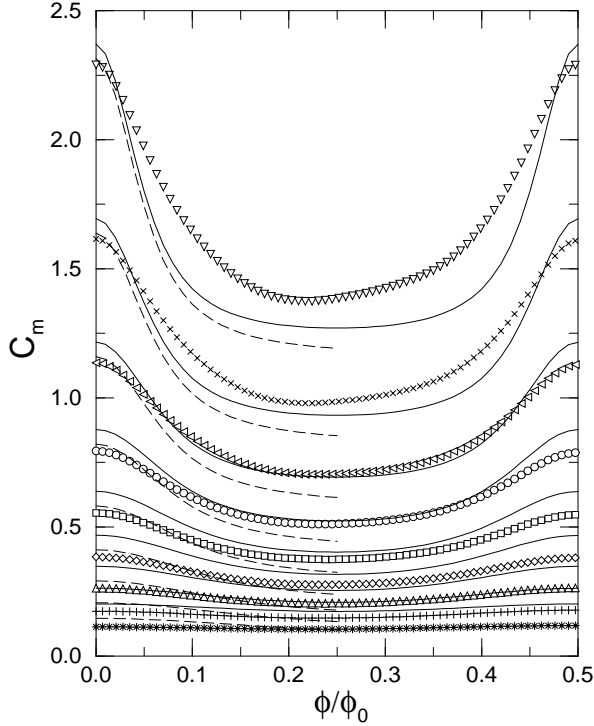,width=\figwidth,angle=0}}
\caption{\label{fig:cm}
  GOE to GUE transition for matrix element correlations according to
  Eq.\ (\protect\ref{eq:cm}) (solid lines) and corresponding results
  from the numerical simulations of the Anderson model (symbols). The
  parameters are the same as in Fig.~\protect\ref{fig:cv} and the
  off-diagonal variance is taken to be $\sigma_{\protect\rm off}^2 =
  0.65$. The dashed lines indicate the $\nu=0$ term of Eq.\ 
  (\protect\ref{eq:cm}) for small $\phi$.}
\end{figure}


\begin{figure}
\centerline{%
  \psfig{file=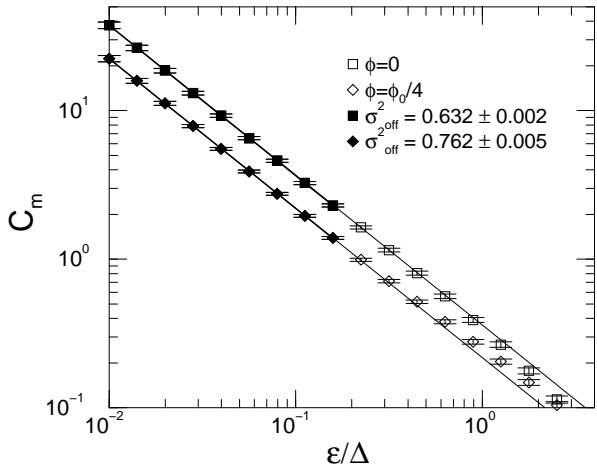,width=\figwidth}}
\caption{\label{fig:cm:small}
  Determination of the off-diagonal variance of the dipole moment
  operator by fitting the values of $C_m$ at small $\epsilon$ (filled
  symbols) to Eq.\ (\protect\ref{eq:cm:small}).  The disorder is
  $W=2.4$.}
\end{figure}


\begin{figure}
  \centerline{\psfig{file=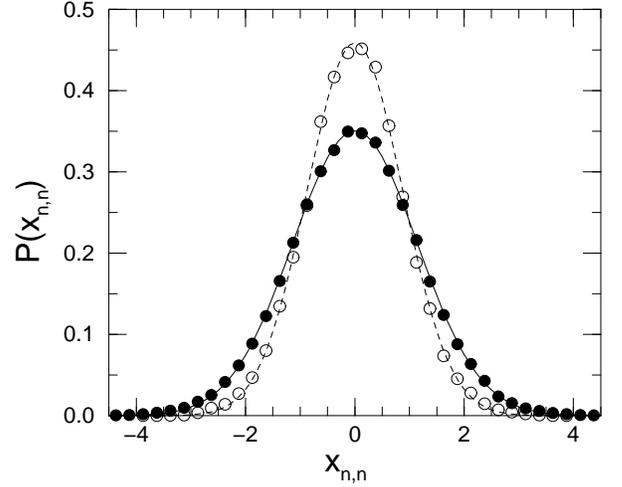,width=\figwidth}}
\caption{\label{fig:mehist:goegue}
  Distribution for diagonal matrix elements of the dipole moment
  operator in units of the lattice constant and in case $W=2.4$,
  $\phi=0$ (filled circles) and averaged over all
  $\phi/\phi_0=0.175,...,0.329$ (open circles) with $\phi_y=0$. The
  lines are fits by Gaussian distributions.  The variance
  $\sigma^2_{\mathrm diag}=2\sigma^2_{\mathrm off}/\beta$ of the data
  is $1.3\pm0.2$ for $\phi=0$ and $0.8\pm0.1$ for
  $\phi/\phi_0=0.175,...,0.329$. The error limits represent the standard
  deviations of the values obtained for different realizations of
  disorder.}
\end{figure}


\begin{figure}
  \centerline{\psfig{file=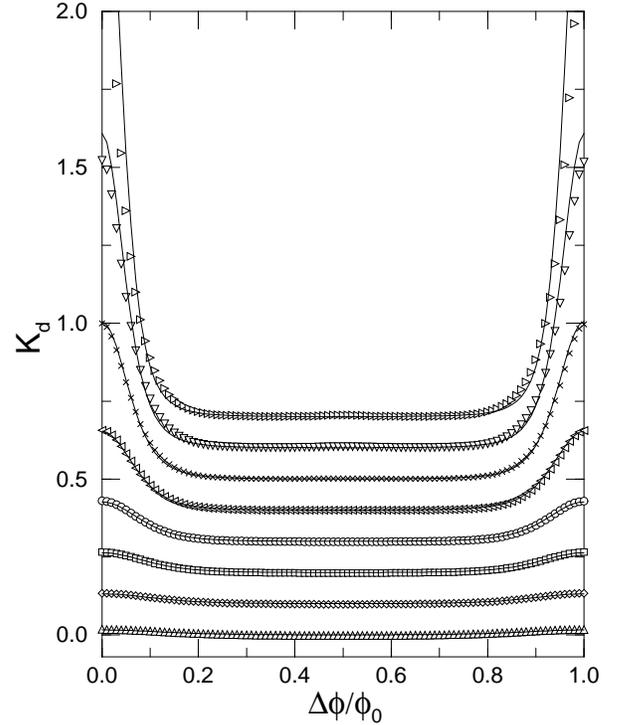,width=\figwidth}}
\caption{\label{fig:kd}
  Parametric correlations of density of states according to Eq.\ 
  (\protect\ref{eq:kd}) (solid lines) compared to the numerical
  results for the Anderson model (symbols) as a function of
  $\Delta\phi$. The parameters are $W=1.7$, $\lambda=2.5$ and $\epsilon
  =$ 0.112 ($\rhd$), 0.158 ($\bigtriangledown$), 0.224 ($\times$),
  0.316 ($\lhd$), 0.447 ($\circ$), 0.631 ($\Box$), 0.891 ($\Diamond$),
  1.26 ($\bigtriangleup$).  The curves have been shifted by
  multiples of $0.1$ for clarity.}
\end{figure}


\begin{figure}
  \centerline{\psfig{file=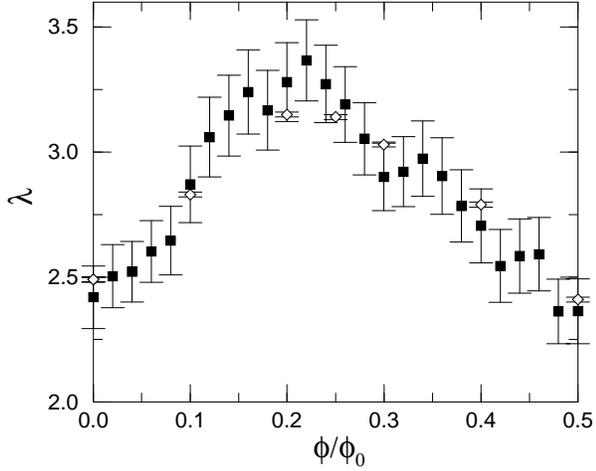,width=\figwidth}}
\caption{\label{fig:diff:gue}
  Parameter $\lambda=2{\cal D}/L^2$ for the Anderson model, determined 
  by fitting
  Eq.\ (\protect\ref{eq:cv:small}) to the data ($\Diamond$) and from
  the variance of the level velocities ($\blacksquare$) with $W=1.7$
  and different flux values in the presence of a transversal flux
  $\phi_y=\phi_0/4$.}
\end{figure}


\begin{figure}
  \centerline{\psfig{file=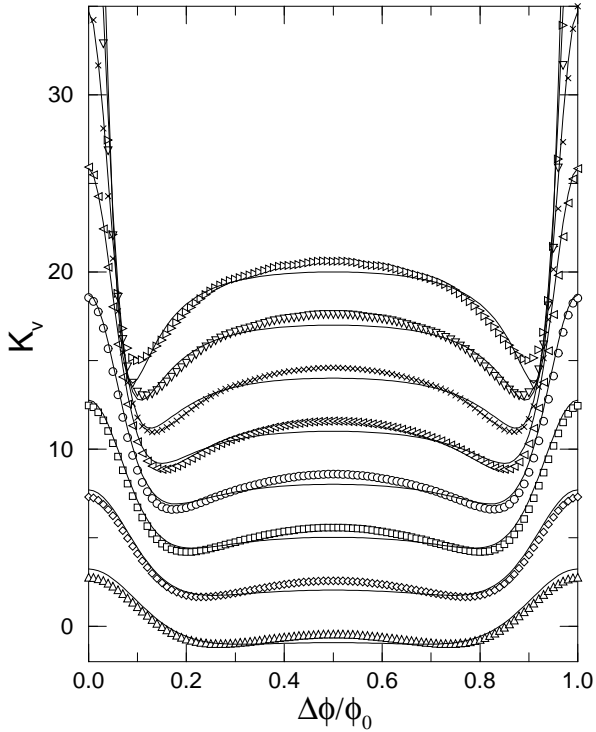,width=\figwidth}}
\caption{\label{fig:kv}
  Parametric correlations of level velocities according to Eq.\ 
  (\protect\ref{eq:kv}) (solid lines) compared to the numerical
  results for the Anderson model (symbols) as a function of
  $\Delta\phi$. The parameters are the same as in
  Fig.~\protect\ref{fig:kd}.  The curves have been shifted by
  multiples of $1$ for clarity.  }
\end{figure}


\begin{figure}
\centerline{%
  \psfig{file=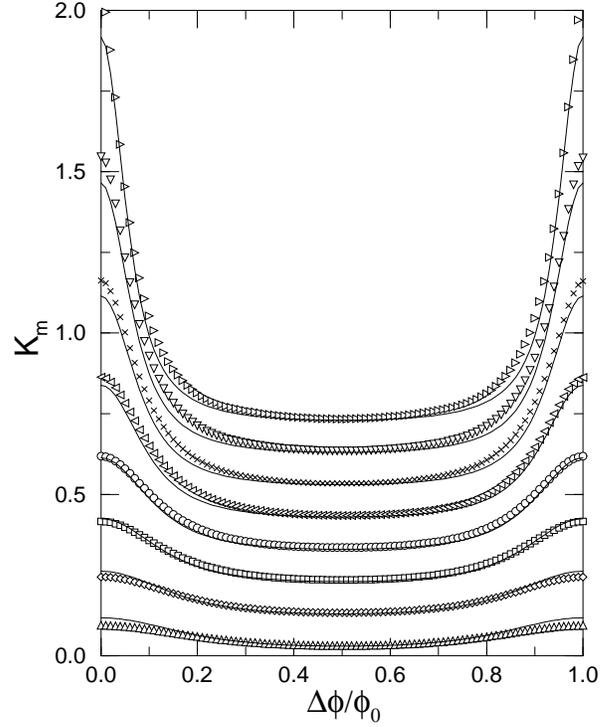,width=\figwidth}}
\caption{\label{fig:km}
  Parametric correlations of matrix elements according to Eq.\ 
  (\protect\ref{eq:km}) (solid lines) compared to the numerical
  results for the Anderson model (symbols) as a function of
  $\Delta\phi$. The parameters are the same as in
  Fig.~\protect\ref{fig:kd} and $\sigma_{\mathrm{off}}^2=0.48$ has
  been used.  The curves have been shifted by
  multiples of $0.1$ for clarity.  }
\end{figure}


\begin{figure}
  \centerline{\psfig{file=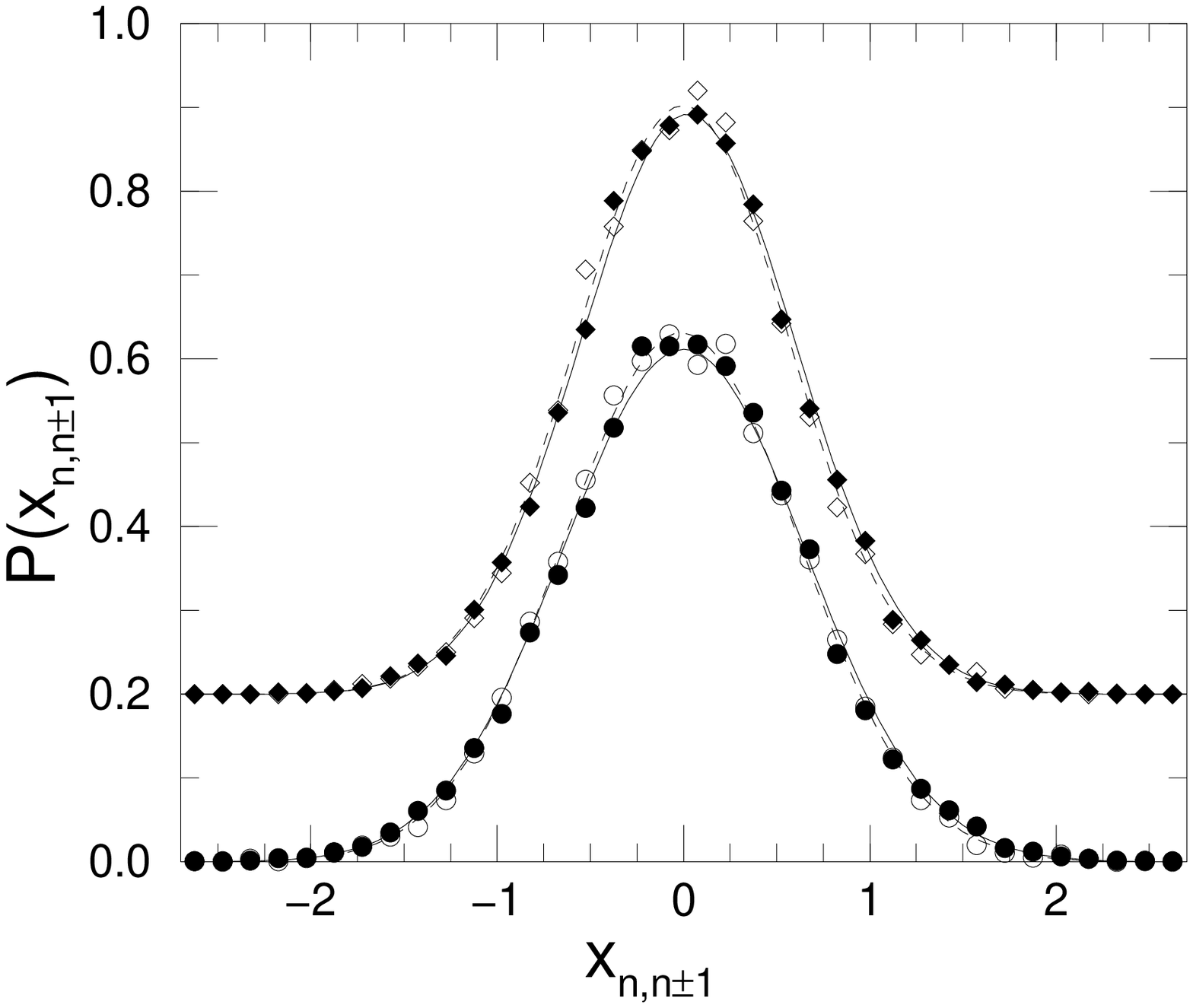,width=\figwidth}}
\caption{\label{fig:off-diag}
  Distribution of real and imaginary parts of the off-diagonal dipole
  matrix elements at flux values $\phi=0$ (filled symbols) and
  $\phi=\phi_0/4$ (open symbols). Additionally, $\phi_y=0$ for $W=2.4$
  ($\circ$) and $\phi_y=\phi_0/4$ for $W=1.7$ ($\diamond$).  The lines
  represent fits by Gaussians. The distributions for $W=1.7$ have been
  shifted by $0.2$ for clarity.}
\end{figure}


\begin{thebibliography}{10}
\bibitem{note} In the infinite volume limit, 
  the semiclassical estimate of
  $\sigma_{\rm off}^2$ for the dipole moment
  operator in a diffusive system diverges. 
  Since we are dealing with a finite system here,
  the definition of $\sigma_{\rm off}^2$ is unproblematic
  in the present case.
\end{thebibliography}
\end{document}